# Comparison of research productivity of Italian and Norwegian professors and universities

**Authors:** Giovanni Abramo[1]*, Dag W. Aksnes[2], Ciriaco Andrea D'Angelo[3]

**Affiliations:**

[1] Laboratory for Studies in Research Evaluation, Institute for System Analysis and Computer Science (IASI-CNR). National Research Council, Rome, Italy
[2] Nordic Institute for Studies in Innovation, Research and Education, Oslo, Norway
[3] University of Rome "Tor Vergata", Dept of Engineering and Management, Rome, Italy

**Abstract**
This is the first ever attempt of application in a country other than Italy of a research efficiency indicator (FSS), to assess and compare the performance of professors and universities, within and between countries. A special attention has been devoted to the presentation of the methodology developed to set up a common field classification scheme of professors, and to overcome the limited availability of comparable input data. Results of the comparison between countries, carried out in the 2011-2015 period, show similar average performances of professors, but noticeable differences in the distributions, whereby Norwegian professors are more concentrated in the tails. Norway shows notable higher performance in Mathematics and Earth and Space Sciences, while Italy in Biomedical Research and Engineering.

______________________
* *corresponding author*

**Acknowledgement**
We wish to thank Gunnar Sivertsen and Lin Zhang for acting as catalysts in this research project

# 1. Introduction

In their article "A farewell to the MNCS and like size-independent indicators" (Abramo & D'Angelo, 2016a), published in a special section of the Journal of Informetrics (2016, vol.10, issue 2), the authors: i) refuted the validity of the "Mean Normalized Citation Score" (MNCS) and all similar per-publication citation indicators as measures of research performance; ii) urged the adoption of efficiency (output to input ratio) indicators, such as the "Fractional Scientific Strength" or FSS (Abramo & D'Angelo, 2014) and highly cited articles per scientist (Abramo & D'Angelo, 2015); and iii) made recommendations on how to expedite the shift to the proposed new paradigm of research performance measurements, in place of the old one of the MNCS and like. The manuscript has elicited a number of responses by eminent scholars in the field.

Most respondents argued that accessing input data could pose formidable problems in most countries. Furthermore, even if input data were accessible, their quality and cross-nation comparability would be doubtful. "Collecting standardized input data requires a high degree of coordination between countries, and it is probably not realistic to expect that this degree of coordination can be achieved" (Waltman, Van Eck, Visser, & Wouters, 2016). Thelwall (2016) feared that the input data provided by research institutions could easily be gamed. Bornmann and Haunschild (2016) warned about possible uncontrollable uses of collected input data: "Since we are not only scientometricians (who appreciate the availability of data), but also researchers ourselves, we should not support such transparent systems which form the 'glass researcher'". For the large part, these are difficulties and apprehensions that Abramo and D'Angelo too can share. The point they raise is whether they are sufficient to induce scientometricians to renounce embarking on a path that can no longer be avoided (Abramo & D'Angelo, 2016b).

What we propose in this work is the first attempt ever to apply the FSS indicator of research performance of professors and universities to a country other than Italy, e.g. Norway. Similarly to Italy, accessing professors' identity and their affiliation in Norway is quite straightforward. The main problem we encountered was the alignment of the research field classification of professors, as the classification schemes in the two countries are rather different. As we show in Section 4, field classification of professors is in fact critical to compare performance, both at the individual and aggregate levels. The comparison of performance of professors from the two countries is not an end per se, but it is also a means to make in each country the performance measures more robust in those fields where the number of observations in either country is low.

The principal aims of the work are to present all the difficulties we have encountered to achieve comparable measurements of performance in the two countries, and to show how we have overcome such difficulties. We have devoted then special care in describing the procedure to operationalize the measurements, and all the limits and assumptions involved. This should make hopeful future replications of the exercise in other countries more straightforward. In this study, we have measured and compared research productivity of Italian and Norwegian professors and universities in the period 2011-2015.

The rest of the paper is structured as follows. In the next section, we will briefly describe the characteristics relevant to our analysis of the higher education systems in the two countries. In Section 3 we present the construction of the dataset, and in Section 4 the method to accomplish research performance comparisons. In Section 5, we will present the results of the assessment. Section 6 will conclude the work with the authors' considerations.

## 2. The higher education systems in Italy and Norway

### 2.1 Italy

The Italian Ministry of Education, universities, and Research (MIUR) designates a total of 98 universities throughout Italy with the authority to issue legally recognized degrees. Of these 30 are small, private, special-focus universities (of which 11 offer only e-learning), and 68 are public and generally multi-disciplinary universities. Six of the whole are *Scuole Superiori* (Schools for Advanced Studies), specifically devoted to highly talented students, with very small faculties and tightly limited enrolment per degree program.

In the overall system, 94.9 per cent of faculty are employed in public universities, and only 0.5 per cent in *Scuole Superiori*. Public universities are largely financed by the government (56 per cent of total income). Until 2009 the core government funding was input oriented (i.e. independent of merit, distributed to universities in a manner intended to equally satisfy the needs of each and all, varying only in respect to size and research disciplines). It was only following the first national research evaluation exercise, conducted in the period 2004-2006, that the MIUR began to assign a minimal share, equivalent to 4 per cent of total income, on the basis of research and teaching assessment. The merit-based share has since increased to 20 per cent.

Despite interventions intended to grant increased autonomy and responsibilities to the universities (Law 168 of 1989),[1] the Italian higher education system is a long-standing, classic example of a public and highly centralized governance structure, with low levels of autonomy at the university level and a very strong role played by the central state.

In keeping with the Humboldtian model, there are no "teaching-only" universities in Italy, as all professors are required to carry out both research and teaching. National legislation includes a provision that each faculty member must provide a minimum of 350 hours per year of instruction (including teaching, preparation to teaching, exams, thesis supervision, etc.). At the close of 2018, there were 54700 faculty members in Italy (full, associate and assistant professors) and a roughly equal number of technical-administrative staff. Salaries are regulated at the central level and are calculated according to role (administrative, technical or professorial), rank within role (e.g. assistant, associate or full professor) and seniority. None of a professor's salary depends on merit.[2] Moreover, as in all Italian public administration, dismissal of unproductive employees is unheard of. All new personnel enter the university system through public

---

[1] This law was intended to grant increased autonomy and responsibility to the universities to establish their own organizational frameworks, including charters and regulations. Subsequently, Law 537 (Article 5) of 1993 and Decree 168 of 1996 provided further changes intended to increase university involvement in overall decision-making on use of resources, and to encourage individual institutions to operate on the market and reach their own economic and financial equilibrium.

[2] In theory, the triennial salary adjustment for inflation depends upon request and evaluation in the period, but it is conceded almost to all.

competitions, and career advancement depends on further public competitions.

New transparency provisions, and timely issue of regulations for the evaluation procedures, are all intended to ensure efficiency in the faculty selection process. In reality, the systemic characteristics – including a historically strong inclination to favoritism, structured absence of responsibility for poor performance by research units, and lack of merit incentive schemes – undermine the credibility of selection procedures for hiring and advancement of university personnel. This lack of credibility is accentuated in the public eye by the high and growing number of legal cases brought by unsuccessful candidates, by continual critical reports in the social and mass media, and by specific studies of systemic problems (Perotti, 2008; Zagaria, 2007).

The overall result is that of a system of universities almost completely undifferentiated for quality and prestige, with the exception of the tiny *Scuole Superiori* and a very small number of private special-focus universities. Thus, also given the widespread knowledge of this context, the system is unable to attract significant numbers of talented foreign faculty, or even students. This is a system where: i) every university has some share of top scientists (TSs), flanked by another share of absolute non-producers (Abramo, Cicero, & D'Angelo, 2013a); ii) 23 per cent of professors alone produce 77 per cent of the overall Italian scientific advancement; iii) this 23 per cent of 'top' faculty is not concentrated in a limited number of universities, but dispersed more or less uniformly among all Italian universities, along with the unproductive academics, so that no single institution reaches the critical mass of excellence necessary to develop as an elite university and to compete internationally (Abramo, Cicero, & D'Angelo, 2012a).

**2.2 Norway**

The Norwegian higher education system consists of several types of institutions, including universities, specialized university colleges and university colleges/universities of applied sciences. Traditionally, the main task of the university colleges has been to offer vocationally oriented education. The universities on the other hand have research as a main task, in addition to providing educational programs from Bachelor to PhD level. Most institutions are state owned. In order to achieve or retain status as various types of higher education institutions, each organization has to be accredited by NOKUT (National Agency for Quality in Education).

This institutional differentiation has been changing in recent years. Both universities and university colleges as well as independent research institutes have merged, and several former university colleges have achieved status as universities. This process was accelerated by a structural reform in 2014 (Ministry of Education and Research, 2014-2015) which further reduced the number of university colleges by merging university colleges together or merging university colleges with universities. As a result of this reform, the distinction between universities and university colleges has blurred.

Currently, there are ten universities in Norway (two traditional universities, two traditional universities which merged with university colleges, one former specialized university institution, and five former university colleges), nine specialized university colleges and 14 university colleges/universities of applied sciences in Norway. The majority of the research in the higher education sector in Norway is carried out at the University of Oslo, the Norwegian University of Science and Technology, the University of Bergen and the Arctic University of Norway. In 2018, these four institutions accounted for 68 per cent of the total scientific and scholarly publication output of the sector.

Since 1995, the academic career structure in Norway has consisted of two types of tracks: a research-oriented and a teaching-oriented (Frølich et al., 2018). The first includes associate professors and full professors as the main types of permanent positions, while the teaching-oriented positions consist of lecturer (*universitets-/høyskolelektor*), senior lecturer (*førstelektor*) and docent (*dosent*). Although the two career systems are applied across types of institutions, the universities have mainly research track positions while the university colleges have a majority of teaching positions.

According to the national regulation, research is neither an individual duty nor a right (Frølich et al., 2018). However, at the traditional universities the common practice has been that the tenured academic staff should spend as much time for teaching as for research. In 2017, the number of tenured academic staff in the higher education sector in Norway was 8,450 (full, associate and assistant professors). In addition, a large number of people are applied in temporary positions as researchers, postdocs, or PhD candidates. Appointments to academic positions are based on public competitions where vacation positions usually are advertised internationally. Since 1993, it has been possible for associate professors to apply for promotion to full professor on the basis of research competence. This is now the most usual way of becoming full professor in Norway (Kyvik, 2015). The promotion system has led to a large increase in the number of full professors, which explains why Norway has a high proportion of full professors compared with many other countries (Frølich et al. 2018). The tenured academic staff at the Norwegian research universities would normally have 50/50 distribution between research and teaching over time. In addition, time is spent on other institutional activities, such as "third mission" technology transfer, and administration. When it comes to salaries, academic rank is the primary determinant, followed by seniority, although additional criteria might be applied (Kyvik, 2010). Traditionally, Norway as a social-democratic society has adopted an egalitarian policy system, which also applies to academic salaries.

**3. Data**

Measuring and comparing research productivity of Italian and Norwegian professors and universities, requires access to the following data: i) professors' name and surname, affiliation, field of research, and academic rank; 2) their research output in the period under observation.

We extract data on the Italian faculty at each university from the database on university personnel, maintained by the MIUR. For each professor this database provides information on their name and surname, gender, affiliation, field classification and academic rank, at close of each year.[3]

We extract Norwegian data from a similar database, the Norwegian Research Personnel Register (providing the official Norwegian R&D statistics, compiled by the Nordic Institute for Studies in Innovation, Research and Education-NIFU). This database contains individual data for all researchers in the higher education sector and institute sector (institutions carrying out R&D which are not part of the higher education and industry sectors) in Norway.

---

[3] http://cercauniversita.cineca.it/php5/docenti/cerca.php, last accessed on 6 June 2019.

For reasons of significance, the analysis is limited to those professors who held formal faculty positions for at least three years (Abramo, D'Angelo, & Cicero, 2012a) over the 2011-2015 period. Furthermore, the dataset is limited to individuals with at least one publication during the time period (non-publishing personnel is not registered in Norwegian databases). As a consequence, productivity measures at the aggregate levels concern only productive professors.

The bibliometric dataset used to assess Italian output is extracted from the Italian Observatory of Public Research (ORP), a database developed and maintained by Abramo and D'Angelo and derived under license from the Clarivate Analytics Web of Science (WoS) Core Collection. Beginning from the raw data of the WoS, and applying a complex algorithm to reconcile the author's affiliation and disambiguation of the true identity of the authors, in ORP each publication is attributed to the university professors that produced it (D'Angelo, Giuffrida, & Abramo, 2011).[4]

Data on publication output of the Norwegian professors is based on a bibliographic database called Cristin (Current Research Information System in Norway), which is a common documentation system for all institutions in the higher education sector, research institutes and hospitals in Norway. Cristin has a complete coverage of the scientific and scholarly publication output of the institutions. Publication data from professional bibliographic data sources (i.e. WoS and Scopus) are imported to the Cristin system, to facilitate the registration of publications by the employees. The publications are indexed as standard bibliographic references, which can be analysed bibliometrically.

In order to obtain the Norwegian dataset, the Cristin publication database has been linked with the Research Personnel Register. Each individual has unique IDs in both databases. However, the IDs are not identical. The linking is based on data on the full name of the professors as well as their institutional affiliations. For a large number of individuals, there is a one-to-one correspondence, and homonyms (different people with identical names) do not represent a problem. In our study we have linked manually professors with identical names, using available data and information.

While the Cristin database contains a complete coverage of the publication output, this study has been limited to the subset of publications which are indexed in the WoS databases, more specifically to the Science Citation Index Expanded, the Social Science Citation index, and the Arts and Humanities Citation Index. As for document types, in the Norwegian system only articles and reviews count as scientific and scholarly publications. As a consequence, the dataset used for comparison contains only the above types of publications. Citations for both Italian and Norwegian publications were counted on 31 October 2018.

As said above, this study is based on WoS data. WoS has a very good coverage of scientific literature within the natural sciences and medicine, but the coverage of technology, the social sciences and humanities has more limitations (Aksnes & Sivertsen, 2019). In particular, this issue relates to the lack of coverage of national journals, non-English literature and books. This means that our analyses are not based on a complete set of the scientific and scholarly publication output of the professors. For reasons of significance, we have excluded analyses of the humanities and some fields of the social sciences, where the coverage of the database has largest limitations. The question would then be whether, in the fields (WoS subject categories, SCs) under observation, the coverage is different across the countries under observation. Although we do not have data to assess this question empirically, previous research suggests that field rather than country is the most important variable when it comes to the coverage of the database, and that publication patterns show fundamental field similarities across all countries (Sivertsen, 2014).

---

[4] The harmonic average of precision and recall (F-measure) of authorships, as disambiguated by the algorithm, is around 97% (2% margin of error, 98% confidence interval).

## 4. Methods

In the following, we present the common field classification scheme developed for professors from both countries, and the research performance indicator.

### 4.1 A common research field classification scheme for professors

Knowing the dominant research field of each professor is a prerequisite of any distortion-free comparative performance assessment. In fact, taking advantage of the Italian professors' field-classification scheme, D'Angelo and Abramo (2015) measured publication rates in 192 research fields, and showed that the intensity of publication remarkably varies not only among disciplines but also among fields within the same discipline. Similar findings have been found for Norwegian researchers (Piro, Aksnes & Rørstad, 2013). The utility of these findings is not limited to solving performance comparison between researchers in different fields, but extends most importantly to the problem of measuring and comparing research performance at the aggregate level of organizational units carrying out research in different fields or with unequal staffing for the respective fields, such as entire departments, schools, universities, or even national higher education systems. In fact, for such unequal organizations, overall measures of performance can only be calculated by beginning from those of their individual researchers, appropriately standardized.

In the Italian university system all academics are classified in one and only one field, named scientific disciplinary sector (SDS), 370 in all. SDSs are grouped into disciplines, named university disciplinary areas (UDAs), 14 in all. In Norway, there is no such a fine-grained classification system. The publications are classified in 4 broader disciplinary areas and 86 fields, and in some analyses this system has been used to field classify the academic staff.

Because of the remarkable difference in the graininess of the two field classification schemes, which makes it formidable to reconcile, we had to figure out a different common classification scheme. We then recurred to the WoS SC classification scheme, and adopted the following procedure to classify each professor in one and only one SC.

First of all, we identified the WoS indexed publications of each professor under observation, over a period of time. We then assigned to each publication the SC or SCs of the hosting journal. Finally, we classified each professor in the most recurrent (dominant) SC in their publication portfolio. To exemplify, we consider the case John Doe, who in the period of observation produced eight articles published in four different journals (*Physical Review B*, *Physical Review E*, *Chemphyschem* and *Physical Review* *letters*). Given the classification of these journals under the WoS system, we have the SC distribution illustrated in Table 1. The eight articles fall in six different SCs (full counting), of which the most recurrent one is UK (Physics, condensed matter), given that it recurs four time.

*Table 1: Publication portfolio of a professor in the dataset*

| SCs | Discipline | No. of publications | WoS_ID |
| --- | --- | --- | --- |
| UK (Physics, condensed matter) | Physics | 4 | 243195800122; 245330200070; 260574500061;251986500011 |
| UF+UR (Physics, fluids & plasmas; Physics, mathematical) | Physics | 2 | 228818200106; 242408800041 |
| EI+UH (Chemistry, physical; Physics, atomic, molecular & chemical) | Chemistry; Physics | 1 | 231971100043 |
| UI (Physics, multidisciplinary) | Physics | 1 | 229700800052 |

A problem arises when the portfolio is limited to one or a few publications or when one observes more than one dominant SC, an event which in turn is more likely when the professor's number of publications is low. Therefore, the larger the observation period of production the better.

To reduce the probability of low number of publications, just for the purpose of identifying the dominant SC of professors, for Italian professors we extended the observation period of production to eleven years: 2006-2016. We adopted a similar procedure for Norwegian professors, extending the observation period to seven years: 2011-2017 (corresponding data for previous years is not available).

Residual cases of professors with more than one dominant SC (14 per cent of Italians and 26 per cent of Norwegians), were solved as follows:

- For the Italian dataset, we measured the frequency distribution of SCs of the 2006-2016 overall publications by all Italian professors in each SDS. Knowing the SDS of each professor, we assigned to him or her the SC (among the dominant ones) with the highest frequency in the publications of professors belonging to the relevant SDS.
- For the Norwegian dataset, we randomly selected one of the dominant SCs attributed to the professor.

Finally, after merging the datasets of the two countries, we included in the final dataset only those SCs (177 in all) with at least: i) one Norwegian and one Italian professor; and ii) ten professors in total.

The final dataset consists of 93 Italian and 6 Norwegian[5] institutions; 34009 Italian and 4327 Norwegian professors. Their distribution per academic rank and discipline[6] is shown in Table 2.

---

[5] Separate assessments have been provided for the 5 largest institutions, only (the University of Oslo, the Norwegian University of Science and Technology, the University of Bergen, the Arctic university of Norway, and the Norwegian University of Life Sciences. The other institutions have been classified together in a single category (named other HE-Institutions). This category consists of other universities and university colleges where the academic staff may devote less time to research than what is the case at the other universities

[6] SCs are grouped in disciplines following a pattern previously published on the website of ISI Journal Citation Reports, but no longer available on the current Clarivate portal. There are no cases in which an SC is assigned to more than one discipline.

*Table 2: Dataset of analysis*

| Discipline | No. of SCs | Italy | | | | Norway | | | |
|---|---|---|---|---|---|---|---|---|---|
| | | Tot. professors | Assistant (%) | Associate (%) | Full (%) | Tot. professors | Assistant (%) | Associate (%) | Full (%) |
| Biology | 28 | 5635 | 34.9 | 37.6 | 27.5 | 736 | 20.7 | 28.0 | 51.4 |
| Biomedical Research | 14 | 3707 | 37.8 | 36.3 | 25.9 | 245 | 19.2 | 30.2 | 50.6 |
| Chemistry | 7 | 1896 | 28.1 | 43.8 | 28.2 | 122 | 14.8 | 29.5 | 55.7 |
| Clinical Medicine | 36 | 7571 | 35.8 | 36.0 | 28.2 | 958 | 10.5 | 31.3 | 58.1 |
| Earth and Space sciences | 11 | 1873 | 29.6 | 41.6 | 28.8 | 413 | 14.8 | 29.1 | 56.2 |
| Economics | 8 | 1848 | 19.0 | 40.4 | 40.6 | 402 | 3.0 | 33.1 | 63.9 |
| Engineering | 34 | 5522 | 26.5 | 40.3 | 33.2 | 426 | 5.2 | 27.2 | 67.6 |
| Mathematics | 6 | 2122 | 22.8 | 40.4 | 36.8 | 183 | 2.2 | 29.0 | 68.9 |
| Physics | 16 | 2918 | 23.8 | 43.3 | 32.9 | 256 | 9.8 | 21.5 | 68.8 |
| Political and social sciences | 11 | 442 | 20.6 | 41.6 | 37.8 | 438 | 6.2 | 33.1 | 60.7 |
| Psychology | 6 | 475 | 31.2 | 39.8 | 29.1 | 148 | 2.7 | 42.6 | 54.7 |
| *Total* | *177* | *34009* | *30.6* | *39.0* | *30.4* | *4327* | *10.9* | *30.1* | *59.0* |

## 4.2 Measuring research performance

Research activity is a production process in which the inputs consist of human resources and other tangible (equipment, scientific instruments, materials etc.) and intangible (accumulated knowledge, social capital, reputation, etc.) resources, and where outputs have a complex character of both tangible (publications, patents, conference presentations, databases, protocols etc.) and intangible nature (tacit knowledge, consulting activity, etc.). Thus, the new-knowledge production function linking outputs to inputs, has a multi-input and multi-output character. The principal indicator of the efficiency of any production system is productivity.

The calculation of productivity requires a few simplifications and assumptions. It has been shown (Moed, 2005) that in the sciences, the prevalent form of codification of research output is publication in scientific journals. As a proxy of total output, in this work we consider only publications (articles and article reviews) indexed in WoS, leaving aside non-indexed publications, patents,[7] databases, other forms of codification of new knowledge, and tacit knowledge.

Most bibliometricians define productivity as the number of publications in the period of observation. Because publications have different values (impact), and resources employed for research are not homogenous across individuals and organizations, we prefer to adopt the definition of productivity extracted from the economic theory of production: the value of output per euro spent in research (i.e. costs of labor and all other production factors, referred to as capital in the following).

This definition recognizes that the publications embedding new knowledge have a different value (impact) on scientific advancement, which bibliometricians generally approximate with citations, or a weighted combination of citations and journal impact factor for short citation time windows (Abramo, D'Angelo, & Felici, 2019; Anfossi,

---

[7] Patents are often followed by publications that describe their content in the scientific arena, so the analysis of publications alone may actually avoid in many cases a potential double counting.

Ciolfi, Costa, Parisi, & Benedetto, 2016). Provided that there is an adequate citation time window (from 3 to 7 years, is our case) the use of citations alone is acceptable (Stegehuis, Litvak, & Waltman, 2015; Stern, 2014; Abramo, Cicero, & D'Angelo, 2011; Levitt & Thelwall, 2011). Because citation behavior varies across fields, we standardize the citations for each publication with respect to the average of the distribution of citations for all publications indexed in the same year and the same SC.[8]

Furthermore, research projects frequently involve a team of scientists, which is registered in the co-authorship of publications. In this case, we account for the fractional contributions of scientists to outputs, which is sometimes further signaled by the position of the authors in the list of authors.

Depending on the objectives of an assessment exercise, it might be appropriate or not accounting for the costs of research (input). It is appropriate, when the objective is to reward best performers (i.e. those who produced more, all resources being equal), or to decide where to allocate funds to maximize returns. It is not appropriate, when the objective is to identify for example the most knowledgeable experts in a field. It is known that accounting for the different costs of capital and labor is a formidable task, because of lack of data at the individual level or, where available, of the assumptions and approximations required.

When comparing productivity at the individual level, we proceed in two ways: one way neglects costs, and the other accounts for them. At the aggregate level, we always account for inputs/costs.

If we neglect costs, the yearly productivity of a professor, termed fractional scientific strength ($FSS_P$, where the subscript $P$ stands for professor), is then defined as:

$$FSS_P = \frac{1}{t}\sum_{i=1}^{N}\frac{c_i}{\bar{c}}f_i \quad [1]$$

where:
$t$ = number of years of work by the professor in period under observation
$N$ = number of publications by the professor in period under observation
$c_i$ = citations received by publication $i$
$\bar{c}$ = average of distribution of citations received for all cited publications in same year and SC of publication $i$
$f_i$ = fractional contribution of researcher to publication $i$.

The fractional contribution equals the inverse of the number of authors in those fields where the practice is to place the authors in simple alphabetical order but assumes different weights in other cases. For Biology, Biomedical research and Clinical medicine, widespread practice in both Italy and Norway is for the authors to indicate the various contributions to the published research by the order of the names in the bylines. For the above disciplines, we thus give different weights to each co-author according to their position in the list of authors and the character of the co-authorship (intra-mural or extra-mural), as suggested in Abramo, D'Angelo, and Rosati (2013).[9]

---

[8] Abramo, Cicero, and D'Angelo (2012b) demonstrated that the average of the distribution of citations received for all cited publications of the same year and SC is the best-performing scaling factor.
[9] If the publication is the outcome of an exclusively intramural collaboration (only one affiliation in the address list), 40% is attributed to both first and last author, and the remaining 20% is divided among all other authors. In contrast, if the publication address list shows extramural collaborations, 30% is attributed to both first and last author; 15% to both second and last but one author; and the remaining 10% is di-

We calculate the productivity of each professor in each SC and express it: i) on a percentile scale of 0-100 (worst to best) for comparison with the performance of all professors of the same SC; and ii) as the ratio to the average productivity of all professors of the same SC with productivity above zero.[10] In general we can exclude that productivity ranking lists may be distorted by variable returns to scale, due to different sizes of universities (Abramo, Cicero, & D'Angelo, 2012b) or by returns to scope of research fields (Abramo, D'Angelo, & Di Costa, 2013).

In the case it is appropriate to account for the cost of inputs, when comparing individuals' productivity or measuring productivity at the aggregate level (SC, discipline, department, school, university), one should account for the different cost of labor and capital available for research, as research units are likely to embed professors with different ranks/salaries and provide them with different resources. Failure to account for the different production factors would result in fact in performance ranking distortions as shown by Abramo, D'Angelo, and Solazzi (2010). The information on individual salaries is unavailable in Italy but the salaries ranges for rank and seniority are published.[11] We had also access to the statistics of average salaries per academic rank in Norway in the period 2012-2014.[12] We have been able then to approximate the salary for each individual as the national average of their academic rank.

As said above, in the Italian university system, salaries are established at the national level and fixed by academic rank and seniority. Thus all professors of the same academic rank and seniority receive the same salary, regardless of their merit and the university that employs them. Also in Norway, the academic rank is the primary salary determinant. Therefore, we do not account for differences in salaries across countries, because a given salary can only "buy" a professor, regardless of her or his merit and country. Instead, we need account for differences in salaries across academic ranks, because higher salaries can buy higher academic ranks, and it has been shown that full professors (more costly to society) are on average more productive than associate professors, and these more productive than assistant professors (Abramo, D'Angelo, & Di Costa, 2011).

As for the cost of capital devoted to research per man/year, in Norway it equals 42693 euro PPP.[13] Corresponding data are not available for Italy. We assume that Italian professors can count on the same amount of resources to conduct research. The assumption then is that capital is equally available to each professor, regardless of academic rank, research field,[14] university, and country.

In the absence of specific individual level data, we assume that each professor devotes the same amount of time (50 per cent) to institutional activities other than

---

vided among all others. The weighting values were assigned following advice from senior Italian professors in the life sciences. The values could be changed to suit different practices in other national contexts.

[10] Abramo, Cicero, and D'Angelo (2013b) demonstrated that the average of the productivity distribution of researchers with productivity above zero is the most effective scaling factor to compare the performance of researchers of different fields.

[11] CINECA-Dalia, https://dalia.cineca.it/php4/inizio_access_cnvsu.php. last accessed on 6 June 2019.

[12] NIFU/ the Norwegian Association of Researchers, https://www.forskerforbundet.no/lonn/lonnsstatistikk/, last accessed on 6 June 2019.

[13] Source: The R&D Statistics Bank, NIFU: http://www.foustatistikkbanken.no/nifu/index.jsp?submode=default&mode=documentation&top=yes&language=en. Last accessed on 6 June 2019.

[14] It is well known that certain research fields require much more capital than others (contrast for example mathematics with physics), but it is correct to normalize capital costs and make them equal to all fields to avoid favouring less consuming fields, when comparing performance.

research (i.e. teaching, technology transfer, administration, etc.), and can count on the same resources, regardless of research field, academic rank, university, and country.

When accounting for costs of labor and capital, assuming that labor and capital concur equally in determining research output, the $FSS_P$ formula becomes the following:

$$FSS_{Pwk} = \frac{1}{\left(\frac{w_r}{2} + k\right)} \cdot \frac{1}{t} \sum_{i=1}^{N} \frac{c_i}{\bar{c}} f_i$$

[2]

where:
$w_r$ = average yearly salary of professor of academic rank $r$, regardless of nation
$k$ = average yearly capital available for research to each professor, regardless of academic rank and nation
$t$ = number of years of work by the professor in period under observation
$N$ = number of publications by the professor in period under observation
$c_i$ = citations received by publication $i$
$\bar{c}$ = average of distribution of citations received for all cited publications in same year and SC of publication $i$
$f_i$ = fractional contribution of professor to publication $i$.

We halve labor costs, because we have assumed that 50 per cent of professors' time is allocated to activities other than research.

As for the cost of labor, $w_r$, Table 3 shows how we have determined it per academic rank, weighting average salaries of professors in the two nations.

*Table 3: Average 2012-2014 salaries in euro PPP (considering an average exchange rate at 12.292 NOK per euro)*

| Academic rank | Italy | | Norway | | Weighted average ($w_r$) |
|---|---|---|---|---|---|
| | No. | Avg salary | No. | Avg salary | |
| Assistant professors | 10,403 | 54,574 | 473 | 55,368 | 54,608 |
| Associate professors | 13,261 | 68,514 | 1,301 | 59,711 | 67,728 |
| Full professors | 10,345 | 102,393 | 2,553 | 74,527 | 96,877 |

*Including tax wedge*

Table 4 summarizes cost of labor, cost of capital and total cost normalization factor per academic rank for both countries. In the following, we will use the total cost normalization factor to report measures of productivity.

*Table 4: Production factors costs (euro) by academic rank*

| Rank | $w_r$ | $k_n$ | $\frac{w_r}{2} + k_n$ | Total cost normalization factor |
|---|---|---|---|---|
| Assistant professors | 54,608 | 42,693 | 69,997 | 1 |
| Associate professors | 67,728 | 42,693 | 76,557 | 1.09 |
| Full professors | 96,877 | 42,693 | 91,132 | 1.30 |

At the aggregate level, the yearly productivity $FSS_A$ for the aggregate unit $A$ is:

$$FSS_A = \frac{1}{RS} \sum_{j=1}^{RS} \frac{FSS_{Pwk_j}}{\overline{FSS_{Pwk}}}$$

[3]

Where:
$RS$ = number of professors in the unit, in the observed period;
$FSS_{Pwk_j}$ = productivity of professor $j$ in the unit;
$\overline{FSS_{Pwk}}$ = productivity of all productive professors in the same SC of professor $j$.

## 5. Results

In this section, we present the results of the performance analyses in the two countries. We start with the individual level comparisons, both accounting and not accounting for costs. We then proceed with the aggregate level analyses, accounting for costs.

### 5.1 Individual level analysis

We have measured the research performance of each professor in their relevant SC. As an example, Table 5 shows the performance of all professors in SC Behavioral sciences. In this field, the $FSS_P$ ranges from 6.04 to 0.004 and the $FSS_{Pwk}$ from 4.65 to 0.003. Thus, at the level of individuals, there are very large variations in the productivity. This is a well-known general bibliometric pattern.

Figures 1 and 2 contrast respectively the quartile and decile performance distributions of professors in both countries. There are some differences in the productivity profile of Norwegian and Italian professors. It is evident that there are larger variations in the distribution of Norwegian professors, which is not surprising considering that the number of observations is much lower for Norway than for Italy. One interesting finding is the higher concentration of Norwegian professors in the top and the bottom tails of the distributions. Thus, the performance of Norwegian professors is more skewed than the Italian.

*Table 5: 2011-2015 research productivity of professors in Behavioral sciences*

| | | | | $FSS_P$ | | $FSS_{Pwk}$ | |
|---|---|---|---|---|---|---|---|
| Prof_ID | Country | Institution | Rank | score | rank | score. | rank |
| 61513 | Italy | University of Trento | Full | 6.040 | 1 | 4.646 | 1 |
| 39439 | Italy | University of Padua | Full | 2.182 | 2 | 1.678 | 2 |
| 39451 | Italy | University of Padua | Full | 2.089 | 3 | 1.607 | 3 |
| 124554 | Norway | UiT- Arctic university | Full | 1.698 | 4 | 1.306 | 6 |
| 18829 | Italy | University of Florence | Assistant | 1.445 | 5 | 1.445 | 4 |
| 209041 | Norway | UiT- Arctic university | Associate | 1.440 | 6 | 1.321 | 5 |
| 30522 | Italy | University of Milan | Associate | 1.342 | 7 | 1.231 | 8 |
| 196762 | Norway | Norwegian University of Life Sciences | Assistant | 1.261 | 8 | 1.261 | 7 |
| 42756 | Italy | University of Parma | Associate | 1.252 | 9 | 1.148 | 9 |
| 61511 | Italy | University of Trento | Associate | 1.156 | 10 | 1.061 | 10 |
| 15388 | Italy | University of Bari | Assistant | 0.874 | 11 | 0.874 | 11 |
| 178405 | Norway | Norwegian University of Life Sciences | Full | 0.849 | 12 | 0.653 | 15 |
| 15387 | Italy | University of Bari | Associate | 0.837 | 13 | 0.768 | 13 |
| 16560 | Italy | University of Cagliari | Associate | 0.833 | 14 | 0.764 | 14 |
| 39888 | Italy | University of Padua | Assistant | 0.822 | 15 | 0.822 | 12 |
| 39443 | Italy | University of Padua | Associate | 0.599 | 16 | 0.550 | 16 |
| 130084 | Norway | Norwegian University of Life Sciences | Associate | 0.471 | 17 | 0.432 | 18 |
| 79667 | Italy | University of Pisa | Assistant | 0.441 | 18 | 0.441 | 17 |
| 126376 | Norway | University of Bergen | Full | 0.400 | 19 | 0.308 | 19 |
| 153593 | Norway | University of Bergen | Associate | 0.202 | 20 | 0.185 | 20 |
| 42839 | Italy | University of Parma | Full | 0.183 | 21 | 0.141 | 22 |
| 191749 | Norway | Norwegian University of Life Sciences | Assistant | 0.177 | 22 | 0.177 | 21 |
| 39351 | Italy | University of Padua | Full | 0.159 | 23 | 0.123 | 24 |
| 54424 | Italy | University of Rome "Tor Vergata" | Assistant | 0.126 | 24 | 0.126 | 23 |
| 57870 | Italy | University of Teramo | Assistant | 0.094 | 25 | 0.094 | 25 |
| 27081 | Italy | University of Milan - Bicocca | Associate | 0.071 | 26 | 0.065 | 26 |
| 142027 | Norway | Norwegian University of Life Sciences | Assistant | 0.048 | 27 | 0.048 | 27 |
| 43908 | Italy | University of Parma | Assistant | 0.028 | 28 | 0.028 | 28 |
| 123852 | Norway | University of Bergen | Full | 0.027 | 29 | 0.020 | 29 |
| 120211 | Norway | Other HE-institutions | Full | 0.004 | 30 | 0.003 | 30 |

*Figure 1: Distribution of 2011-2015 research productivity quartiles of Italian (dark bars) and Norwegian (light bars) professors not accounting for cost (left panel) and accounting for cost (right panel)*

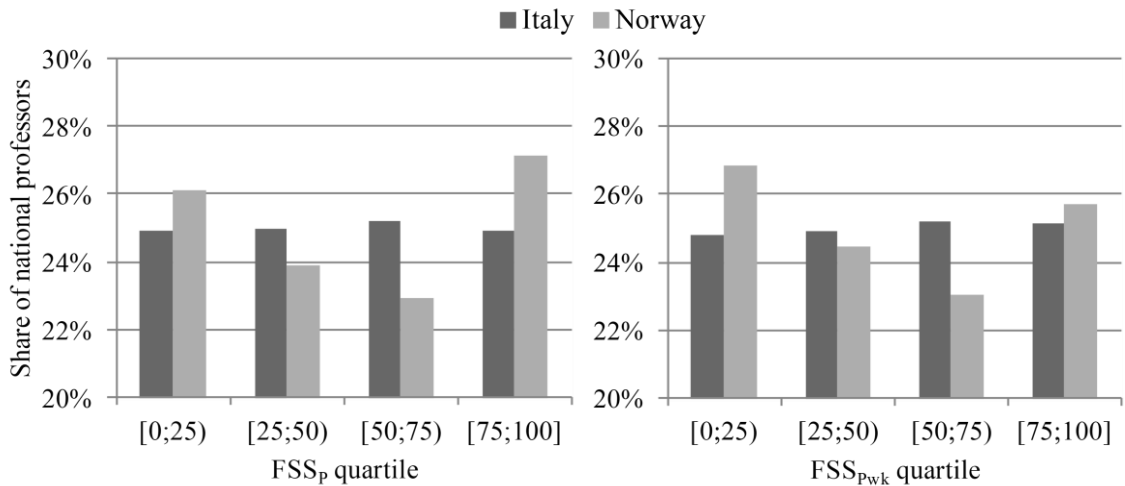

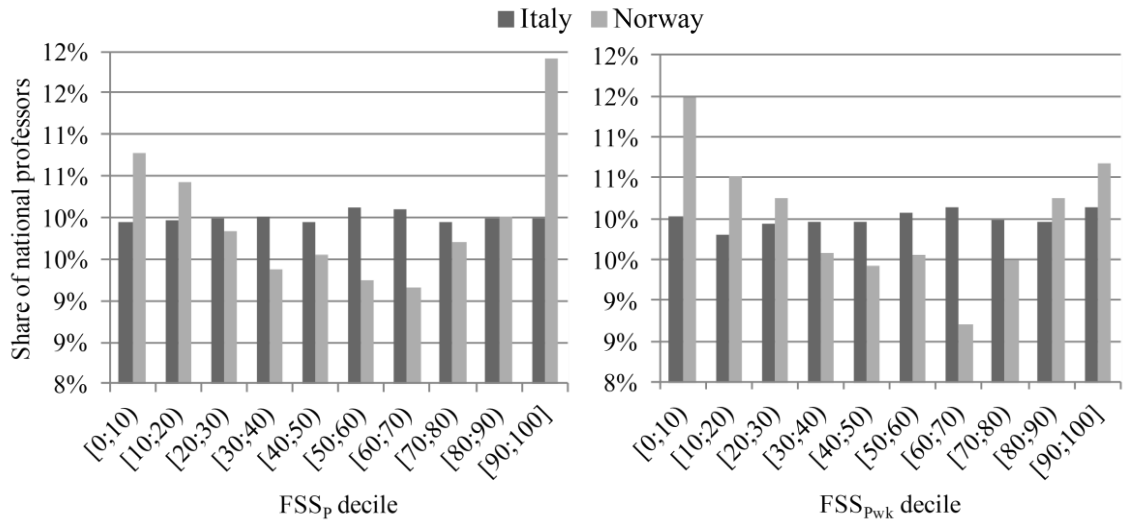

*Figure 2: Distribution of 2011-2015 research productivity deciles of Italian (dark bars) and Norwegian (light bars) professors accounting (left panel) and not accounting (right panel) for cost*

We further delve into the top decile performers at SC and discipline level, and measure in each SC, for each country, the total number of top 1 per cent scientists (TS_1%), top 5 per cent scientists (TS_5%), and top 10 per cent scientists (TS_10%). It is possible then to divide the number of TSs by the total faculty in the relevant SC; or to sum the TSs in the SCs of a discipline and then divide the sum by the total faculty in the discipline. The share of TSs in an SC or discipline is an indicator of excellence that can complement the indicator of productivity to rank universities (Abramo, D'Angelo, & Soldatenkova, 2016), and countries. Incidentally, it has been shown that TSs contribute more than unproductive researchers[15] to the overall research performance of universities (Abramo, Cicero, & D'Angelo, 2013a), and logic would have it that in competitive higher education systems, top research universities would achieve this status from a concentration of TSs.

The Tables 6 and 7 contrast the share of TSs by discipline, respectively not accounting and accounting for cost. When costs are not accounted for, we observe that the share of TSs is higher than expected in the majority of the cells. This holds both for Italy and Norway. For Italy there are two disciplines, Mathematics and Earth and Space Sciences, where the share of TSs is lower than expected in all TS categories, and five disciplines in which it is higher (Biology, Biomedical Research, Clinical medicine, Engineering and Economics). In the four remaining disciplines the pattern is mixed. For Norway, there are eight disciplines with higher TS shares than expected in all TS categories (Mathematics, Physics, Earth and Space Sciences, Clinical Medicine, Psychology, Engineering, Political and social sciences, and Economics). At the overall level, both countries perform on par or above average. As expected from the performance distributions, Norway tends to have higher scores than Italy but when accounting for cost, given the high share of full professors (59% in Norway against 30% in Italy), the differences are somewhat reduced.

---

[15] We cannot measure the share of unproductive professors because the Cristin database does not index them.

*Table 6: Share of top scientists (%) by $FSS_P$ (not accounting for cost)*

|  | TS_1% | | TS_5% | | TS_10% | |
|---|---|---|---|---|---|---|
| Discipline | Italy | Norway | Italy | Norway | Italy | Norway |
| Mathematics | 0.99 | 3.28 | 4.76 | 8.74 | 9.47 | 17.49 |
| Physics | 1.27 | 1.95 | 5.00 | 8.20 | 9.90 | 14.06 |
| Chemistry | 1.16 | 0.82 | 5.06 | 6.56 | 9.81 | 15.57 |
| Earth and Space Sciences | 0.75 | 2.91 | 4.32 | 9.44 | 8.86 | 16.46 |
| Biology | 1.19 | 1.49 | 5.27 | 4.89 | 10.28 | 9.51 |
| Biomedical Research | 1.21 | 1.22 | 5.18 | 4.49 | 10.25 | 8.57 |
| Clinical Medicine | 1.18 | 1.57 | 5.14 | 5.53 | 10.04 | 11.27 |
| Psychology | 1.05 | 2.70 | 4.84 | 7.43 | 9.05 | 14.86 |
| Engineering | 1.25 | 2.11 | 5.16 | 6.10 | 10.20 | 11.03 |
| Political and social sciences | 2.04 | 1.37 | 5.43 | 5.48 | 9.50 | 11.64 |
| Economics | 1.19 | 1.49 | 5.03 | 5.72 | 10.12 | 10.45 |
| Overall | 1.18 | 1.80 | 5.08 | 6.19 | 9.99 | 11.93 |

*Table 7: Share of top scientists (%) by $FSS_{Pwk}$ (accounting for cost)*

|  | TS_1% | | TS_5% | | TS_10% | |
|---|---|---|---|---|---|---|
| Discipline | Italy | Norway | Italy | Norway | Italy | Norway |
| Mathematics | 0.99 | 3.28 | 4.85 | 7.65 | 9.66 | 15.30 |
| Physics | 1.30 | 1.56 | 5.21 | 5.86 | 10.25 | 10.16 |
| Chemistry | 1.16 | 0.82 | 5.17 | 4.92 | 9.92 | 13.93 |
| Earth and Space Sciences | 0.75 | 2.91 | 4.48 | 8.72 | 9.02 | 15.74 |
| Biology | 1.22 | 1.22 | 5.31 | 4.62 | 10.40 | 8.56 |
| Biomedical Research | 1.24 | 0.82 | 5.23 | 3.67 | 10.30 | 7.35 |
| Clinical Medicine | 1.19 | 1.46 | 5.18 | 5.22 | 10.14 | 10.44 |
| Psychology | 1.05 | 2.70 | 5.26 | 6.08 | 9.89 | 12.16 |
| Engineering | 1.25 | 2.11 | 5.22 | 5.40 | 10.30 | 9.62 |
| Political and social sciences | 2.26 | 1.14 | 5.43 | 5.48 | 10.18 | 10.96 |
| Economics | 1.24 | 1.24 | 5.03 | 5.72 | 10.34 | 9.45 |
| Overall | 1.20 | 1.64 | 5.15 | 5.62 | 10.14 | 10.68 |

### 5.2 Aggregate level analysis

As said above, at the aggregate level we always account for costs. In the following, we provide examples of results at the level of overall institution, discipline, and SC. As noted in the Methods section, the dataset consists of professors affiliated with 93 Italian and 6 Norwegian universities (including one for other HE-institutions). The performance of these institutions varies significantly overall and across fields.

Table 8 shows the top 15 and bottom 15 institutions at overall level. At the level of institutions the $FSS_A$ score varies from 0.624 to 3.134. Thus, there are very large institutional differences in productivity. University Vita - Salute San Raffaele is the institution with the highest $FSS_A$ score overall. It is a recently founded private university, specialized in medicine. Three public schools for advanced studies follow in the ranking. Among the top 15 institutions, we find three Norwegian, the first being the University of Oslo (11th in the ranking list).

Table 9 contrasts the research productivity of the two countries at discipline and overall level. Overall, there are hardly any differences at all in the average research productivity across the two countries. The $FSS_A$ indicator is 0.990 for Italy and 0.991 for Norway. However, at the level of disciplines there are larger differences, showing different specializations. While the $FSS_A$ indicator ranges from 0.907 to 1.008 for Italy, the Norwegian figures show larger variations, varying from 0.878 to 1.264. This is related to the fact that Norwegian professors account only for 11 per cent of the total population. By discipline, the Italian scores are higher than the Norwegian in seven out of eleven disciplines. The reason why the $FSS_A$ scores of the two countries still are equal at the overall level is that Norwegian score is highest in the largest field, Clinical Medicine, and that the Norway scores in two fields (Mathematics and Earth and Space Sciences) are much higher than the Italian.

*Table 8: Top 15 and bottom 15 institutions by 2011-2015 research productivity at overall level*

| Institution | Country | Obs | $FSS_A$ |
|---|---|---|---|
| University Vita - Salute San Raffaele | Italy | 87 | 3.134 |
| Scuola Normale Superiore | Italy | 39 | 2.487 |
| Scuola Internazionale Superiore di Studi Avanzati | Italy | 68 | 2.480 |
| Scuola Superiore S.Anna | Italy | 84 | 1.758 |
| University Luigi Bocconi | Italy | 157 | 1.653 |
| University of Trento | Italy | 325 | 1.480 |
| Free University of Bolzano | Italy | 74 | 1.433 |
| University of Padua | Italy | 1560 | 1.332 |
| University "Campus Bio-medico" | Italy | 99 | 1.299 |
| Polytechnic of Milan | Italy | 920 | 1.206 |
| University of Oslo | Norway | 906 | 1.184 |
| University of Milan | Italy | 1478 | 1.184 |
| University of Verona | Italy | 415 | 1.142 |
| Norwegian University of Life Sciences | Norway | 367 | 1.138 |
| Norwegian University of Science and Technology | Norway | 784 | 1.107 |
| … | | | |
| University of Sannio | Italy | 132 | 0.816 |
| University of Trieste | Italy | 400 | 0.814 |
| University of Palermo | Italy | 1012 | 0.803 |
| University of Parma | Italy | 630 | 0.799 |
| University of Aquila | Italy | 391 | 0.786 |
| University of Gabriele D'Annunzio | Italy | 410 | 0.780 |
| University of Cagliari | Italy | 630 | 0.778 |
| University of Siena | Italy | 453 | 0.764 |
| University of Naples "Parthenope" | Italy | 160 | 0.754 |
| Other HE-institutions | Norway | 1267 | 0.745 |
| University of Basilicata | Italy | 218 | 0.731 |
| University of Camerino | Italy | 203 | 0.703 |
| University of Sassari | Italy | 405 | 0.691 |
| University of Molise | Italy | 130 | 0.690 |
| University of Macerata | Italy | 30 | 0.624 |

*Table 9: 2011-2015 research productivity of countries, by discipline and overall*

| | Italy | | Norway | |
|---|---|---|---|---|
| Discipline | Obs | $FSS_A$ | Obs | $FSS_A$ |
| Mathematics | 2122 | 0.942 | 183 | 1.264 |
| Physics | 2918 | 0.998 | 256 | 0.966 |
| Chemistry | 1896 | 0.999 | 122 | 0.977 |
| Earth and Space Sciences | 1873 | 0.937 | 413 | 1.256 |
| Biology | 5635 | 1.008 | 736 | 0.903 |
| Biomedical Research | 3707 | 1.005 | 245 | 0.885 |
| Clinical Medicine | 7571 | 0.990 | 958 | 1.030 |
| Psychology | 475 | 0.976 | 148 | 1.043 |
| Engineering | 5522 | 0.999 | 426 | 0.878 |
| Political and social sciences | 442 | 0.907 | 438 | 0.891 |
| Economics | 1848 | 0.976 | 402 | 0.953 |
| Overall | 34009 | 0.990 | 4327 | 0.991 |

As an example of institution ranking at discipline level, Table 10 shows the top 10 and bottom 10 institutions in the discipline of Engineering.

*Table 10: Top 10 and bottom 10 institutions by 2011-2015 research productivity in the discipline of Engineering*

| Institution | Country | Obs | $FSS_A$ |
|---|---|---|---|
| Norwegian University of Life Sciences | Norway | 10 | 2.374 |
| University of Trento | Italy | 93 | 2.108 |
| Scuola Superiore S.Anna | Italy | 32 | 2.070 |
| University of Insubria | Italy | 17 | 1.971 |
| University of Salerno | Italy | 137 | 1.475 |
| University of Padua | Italy | 211 | 1.365 |
| University of Messina | Italy | 57 | 1.317 |
| University of Catanzaro "Magna Grecia" | Italy | 11 | 1.291 |
| University "Campus Bio-medico" | Italy | 20 | 1.271 |
| University of Bergen | Norway | 34 | 1.259 |
| … | | | |
| UiT- Arctic university | Norway | 14 | 0.747 |
| University of Palermo | Italy | 144 | 0.734 |
| University of Enna | Italy | 11 | 0.730 |
| University of Bergamo | Italy | 41 | 0.661 |
| University of Turin | Italy | 72 | 0.605 |
| University of Oslo | Norway | 54 | 0.598 |
| Tuscia - University of Viterbo | Italy | 10 | 0.572 |
| University of Trieste | Italy | 49 | 0.526 |
| Other HE-institutions | Norway | 126 | 0.492 |
| University "G. D'Annunzio" of Chieti-Pescara | Italy | 17 | 0.431 |

Table 11 shows in each discipline the number of SCs in which Norway outperforms Italy, which only in part reflects the results at discipline level, signalling further different specializations at SC level. For example, while Norway outperforms Italy in Clinical Medicine (see row 8 of Table 9), it does so in a lower number of SCs within the discipline (17 out of 36).

*Table 11: Number of SCs where 2011-2015 research productivity of Norway is higher than Italy, by discipline and overall*

| Discipline | No. of SCs | No. of SCs where Norway outperforms Italy |
|---|---|---|
| Mathematics | 6 | 3 (50.0%) |
| Physics | 16 | 8 (50.0%) |
| Chemistry | 7 | 3 (42.9%) |
| Earth and Space Sciences | 11 | 8 (72.7%) |
| Biology | 28 | 7 (25.0%) |
| Biomedical Research | 14 | 5 (35.7%) |
| Clinical Medicine | 36 | 17 (47.2%) |
| Psychology | 6 | 2 (33.3%) |
| Engineering | 34 | 11 (32.4%) |
| Political and social sciences | 11 | 5 (45.5%) |
| Economics | 8 | 2 (25.0%) |
| Overall | 177 | 71 (40.1%) |

Table 12 shows the ten SCs where the productivity gap of Norway vs Italy is the highest, and the ten SCs where the productivity gap of Italy vs Norway is the highest.

*Table 12: Ten SCs with the highest productivity gap in favor of Norway, and the same for Italy*

|  | Italy | | Norway | | |
|---|---|---|---|---|---|
| SC | Obs | $FSS_A$ | Obs | $FSS_A$ | Δ |
| Mathematics, interdisciplinary applications | 46 | 0.516 | 6 | 4.379 | -3.863 |
| Remote sensing | 102 | 0.847 | 8 | 2.947 | -2.100 |
| Statistics & probability | 390 | 0.884 | 29 | 2.217 | -1.333 |
| History & philosophy of science | 62 | 0.614 | 10 | 1.895 | -1.281 |
| Pediatrics | 182 | 0.887 | 17 | 2.149 | -1.262 |
| Metallurgy & metallurgical engineering | 47 | 0.828 | 9 | 1.785 | -0.957 |
| Substance abuse | 9 | 0.480 | 10 | 1.368 | -0.889 |
| Sport sciences | 127 | 0.763 | 51 | 1.572 | -0.809 |
| Physics, condensed matter | 203 | 0.959 | 10 | 1.723 | -0.763 |
| Physics, nuclear | 92 | 0.934 | 9 | 1.671 | -0.737 |
| … | | | | | |
| Optics | 238 | 1.050 | 19 | 0.320 | 0.730 |
| Urban studies | 26 | 1.047 | 6 | 0.296 | 0.751 |
| Engineering, environmental | 29 | 1.129 | 6 | 0.376 | 0.753 |
| Biology | 11 | 1.236 | 5 | 0.481 | 0.755 |
| Physics, multidisciplinary | 157 | 1.038 | 11 | 0.274 | 0.764 |
| Ophthalmology | 186 | 1.018 | 8 | 0.210 | 0.807 |
| Rehabilitation | 47 | 1.414 | 38 | 0.488 | 0.925 |
| Biodiversity conservation | 6 | 1.569 | 9 | 0.621 | 0.948 |
| Computer science, interdisciplinary applications | 44 | 1.122 | 10 | 0.163 | 0.959 |
| Information science & library science | 13 | 1.790 | 24 | 0.530 | 1.260 |

The analyses shown in Tables 11 and 12, identify relative strengths and weaknesses respectively at discipline and SC levels, and are able to inform research policies and initiatives.

Finally, as an example of institution ranking at SC level, Table 13 shows the rankings of institutions in the SC Geosciences, multidisciplinary.

*Table 13: 2011-2015 research productivity of all institutions in the SC Geosciences, multidisciplinary*

| Institution | Country | Obs | $FSS_A$ |
|---|---|---|---|
| University of Florence | Italy | 20 | 2.210 |
| University of Milan - Bicocca | Italy | 13 | 1.800 |
| University of Oslo | Norway | 28 | 1.709 |
| University of Bergen | Norway | 32 | 1.597 |
| UiT- Arctic University of Norway | Norway | 13 | 1.497 |
| University of Padua | Italy | 31 | 1.496 |
| University of Calabria | Italy | 16 | 1.341 |
| University of Rome "La Sapienza" | Italy | 29 | 1.234 |
| University of Milan | Italy | 14 | 1.090 |
| University of Rome - Roma Tre | Italy | 14 | 1.017 |
| University of "Basilicata" | Italy | 12 | 0.970 |
| University of Naples "Federico II" | Italy | 26 | 0.842 |
| University of Bologna | Italy | 26 | 0.820 |
| Other HE-institutions | Norway | 23 | 0.806 |
| University of Catania | Italy | 15 | 0.779 |
| University of Urbino "Carlo Bo" | Italy | 10 | 0.773 |
| University of Ferrara | Italy | 12 | 0.768 |
| University of Pisa | Italy | 20 | 0.736 |
| University of Parma | Italy | 14 | 0.735 |
| University of Siena | Italy | 11 | 0.691 |
| University of Palermo | Italy | 14 | 0.675 |
| University of Bari | Italy | 20 | 0.655 |
| University of Trieste | Italy | 10 | 0.635 |
| Norwegian University of Science and Technology | Norway | 11 | 0.586 |
| University of Turin | Italy | 23 | 0.534 |
| University "G. D'Annunzio" of Chieti-Pescara | Italy | 13 | 0.506 |
| Polytechnic of Turin | Italy | 11 | 0.363 |

## 6. Discussion and conclusions

The main aim of this study was to verify the feasibility of applying a research efficiency (output-to-input) indicator, namely the FSS (Abramo & D'Angelo, 2014), to assess the standing of higher education systems of countries other than Italy. Thanks to the bibliographic database Cristin, indexing among others the WoS publications of all professors in Norwegian universities, we have been able to assess the individual and aggregate performances of Norwegian universities and compare them to Italy.

The main contribution of the study, in view of hopeful future extensions to other countries, is the methodology developed to overcome the limitation and comparability of input data, and differences in professor field classification schemes. All solutions and assumptions adopted respond to the criterion of making performance comparisons as equitable as possible. The availability of the cost of capital per Norwegian professor made it possible to improve the measurement of individual productivity, that has been so far limited to accounting for the cost of labor only (Abramo & D'Angelo, 2014). Despite the arguments that have been put forward that the FSS indicator is not applicable in cross-national analyses due to problems with standardizing input data (Waltman, Van Eck, Visser, & Wouters (2016), our study has shown that it is possible to apply the indicator in such contexts, provided that necessary harmonization of the data sets is carried out. Hoping that other countries will follow suit, the moral of the study can be expressed with Seneca's words '*Non quia difficilia sunt non audemus, sed quia non audemus difficilia sunt*', or '*It is not because things are difficult that we do not try; it is because we do not try that things are difficult*'.

One interesting finding of this comparison is that overall there are hardly any differences at all in the average research productivity of Italian and Norwegian professors. Thus, despite representing countries which differ considerably along such dimensions as size, research profile and organisation of the research systems, the overall performance measured by the $FSS_A$ indicator is almost equal. Another interesting finding is the higher concentration of Norwegian professors in the top and the bottom tails of the productivity distribution.

Especially for small-sized countries in research terms, the importance of comparing performance with large-sized countries is twofold. On the one hand, the comparison allows a strategic analysis of strengths and weaknesses at field and discipline levels, as compared to other countries, informing then research policies, organizational strategies, fund allocation, incentive systems, etc. In the case in point, results show the notable productivity advantage of Norway in Mathematics and Earth and Space Sciences, and that of Italy in Biomedical Research and Engineering. At field level, Norway notably outperforms Italy in Mathematics, interdisciplinary applications; Remote sensing; and Statistics & probability. The opposite is true in Information science & library science; Computer science, interdisciplinary applications; and Biodiversity conservation.

On the other hand, the comparison with a larger country contributes to make the assessment more robust in those fields where the number of observations reveals too low, which is frequently the case in small-sized countries.

In interpreting the results of the performance analysis, it should be kept in mind that all the usual limits, caveats, assumptions and qualifications of evaluative scientometrics apply, in particular: i) publications as not representative of all knowledge produced; ii) bibliometric repertories applied (WoS) do not cover all publications; and iii) citations are not always certification of real use and representative of all use. Furthermore, results are sensitive to the classification schemes adopted for both publications and professors. Finally, the limited availability of comparable input data required the adoption of few assumptions. Subject to the availability of input data, the present study can be replicated to include other countries or single research institutions of other countries.

Comparative research assessment data provides governments, universities, industry and prospective students with valuable information about research performance in a country's higher education institutions. A rigorous and fine-grained, disciplined based information about research performance that is not readily available through other means. Such data allows research managers and investors to identify and assess performance in research and opportunities for further development; assists institutions with their strategic planning, decision making and their research promotional activities in the country and internationally; helps reduce the information asymmetry between research suppliers and demand, i.e. students, industry and other stakeholders, making it possible to optimize their choices and inducing a continuous improvement on the supply side to be "chosen". Moreover, monitoring research performance and using results, assures taxpayers that their investment in research is well spent.

The results of the current comparison offer a number of stimuli to further delve into the performance analyses. It emerged in fact a different distribution of individual performance between the two countries. Future research might investigate the dispersion of research performance within and between universities. The comparison could reveal different competitive intensity in the two higher education systems.

Furthermore, the single components of the FSS indicator could be compared, namely the intensity of publication, the average impact per publication, and the collaboration behavior.

The two countries present also a quite different history of female emancipation, and attitudes concerning the role of women in the family, in the workplace, and in the society in general. It would be interesting then to explore whether gender representation across academic ranks, and differences in performance occur to the same extent in the two countries.